\documentstyle[aps,preprint]{revtex}
\begin{document}
\tightenlines

\title{Pair contact process with a particle source}

\author {Ronald Dickman$^{\dagger,1}$, Wilson R. M. Rab\^elo$^{\dagger}$
and G\'eza \'Odor$^{\ddagger,2}$}
\address{
$^{\dagger}$Departamento de F\'\i sica, ICEx,
Universidade Federal de Minas Gerais, Caixa Postal 702,
30161-970, Belo Horizonte - MG, Brasil\\
$^{\ddagger}$Research Institute for Technical Physics and Materials Science,
P.O. Box 49, H-1525 Budapest, Hungary}
\date{\today}

\maketitle
\begin{abstract}
We study the phase diagram and critical behavior of the one-dimensional
pair contact process (PCP) with a particle source using cluster approximations 
and extensive simulations.  The source creates isolated
particles only, not pairs, and so couples not to the order parameter 
(the pair density) but to a non-ordering field, whose state influences 
the evolution of the order parameter.  While the critical point $p_c$ shows a 
singular dependence on the source intensity,  the critical exponents 
appear to be unaffected by the presence of the source, except possibly
for a small change in $\beta$.
In the course of our study we obtain high-precision values for the
critical exponents of the standard PCP, confirming directed-percolation-like
scaling.
\vspace {0.6cm}

\end{abstract}


\newpage
\section{Introduction}

Critical phenomena at absorbing-state phase transitions are of 
longstanding interest in statistical physics \cite{privbook,marro,hinr00},
being found, for example, in models of epidemics \cite{harris},
catalytic kinetics \cite{zgb,evans}, surface growth \cite{alon}, 
self-organized criticality \cite{gz,maslov,dvz,vdmz}, and turbulence 
\cite{pomeau,chate,bohr}.
The transition between active and absorbing states arises out of a conflict between two
opposing processes (e.g., creation and annihilation); when continuous
(as is often the case), it falls 
generically in the DP universality class \cite{janssen,gr1,glb}.
When two or more absorbing states exist and are connected by a symmetry
operation, as in branching and annihilating random walks,
a new kind of critical behavior appears
\cite{gkt,baw1,iwanbaw,cardybaw}.  
 
Unusual critical behavior also 
arises in models that can become trapped in one of an infinite
number of absorbing configurations (INAC).
(More precisely, the number of absorbing configurations grows exponentially
with the system size. There is no special symmetry 
linking the different absorbing  configurations.)
Models of this sort were introduced to describe surface 
catalysis \cite{benav,albdd}; their critical properties 
have been studied in detail 
by various workers \cite{ijnoco,pcp1,pcp2,mendes,inas,snr,mgd,gcr}.
In one dimension, the pair contact process (PCP) \cite{pcp1}, and other 
models with INAC exhibit static critical behavior in the DP class 
\cite{pcp2,rdjaff}, but the critical exponents $\delta$ and $\eta$,
associated with the spread
of activity from a localized seed, vary 
continuously with the particle density $\phi$ in the 
environment \cite{pcp2,mendes,gcr,lopez,odor88,lipowski}.
[These exponents are defined via the asymptotic 
($t \!\rightarrow \!\infty$) power laws:
survival probability $P(t) \sim t^{-\delta}$, 
and mean activity $n(t) \sim t^{\eta}$; note also
that the mean-square distance of activity from the seed, $R^2 (t) \sim t^z$.]
This anomalous aspect of critical spreading for INAC can be
traced to a long memory in the dynamics of the order parameter, $\rho$,
arising from a coupling 
to an auxiliary field (the local particle density, $\phi$), 
that remains frozen in regions where
$\rho \!=\! 0$ \cite{inas,mgd,gcr}.  
Theoretical understanding of models with INAC remains incomplete.
Mu\~noz et al. proposed a continuum theory 
for models with INAC, and showed that the static critical behavior is that of DP \cite{inas}.
Formally eliminating the auxiliary field, they obtained a closed equation for the order parameter,
in which a memory term appears; simulations of this theory
also show nonuniversal exponents \cite{lopez}.  
While a theoretical analysis of variable spreading exponents 
in the PCP is lacking, an analogous phenomenon has been analyzed exactly
in the simpler
cases of a random walk with a movable partial reflector \cite{rwmpr},
or of compact DP confined to a parabola \cite{odor00}.

In previous studies of the PCP, and, indeed, of all nonequilibrium 
models possessing an auxiliary or non-ordering field $\phi$, the latter 
has been allowed to relax to its stationary
value \cite{ttpnote}. Only the effect of varying $\phi$ in the initial condition
has been explored. Given the surprising results of these studies, 
it is of interest to investigate the consequences of changing the 
non-ordering field in the {\it stationary state} as well.  To this end, we 
introduce an external field $h$
that couples to $\phi$, but not to the order parameter itself.  In 
concrete terms, $h$ is the rate (per site) of attempted insertions of 
{\it isolated} particles. The source
may only insert a particle at a vacant site, both of whose neighbors 
are also vacant; in this way no pairs are created, and the absorbing 
nature of configurations devoid of pairs is maintained.  In this work we 
examine the effect of this perturbation of the phase diagram and critical
behavior of the PCP, using cluster approximations and extensive 
simulations.

The balance of this paper is devoted to
defining the model (Sec. II);
a discussion of cluster
approximations (Sec. III);
and analysis of simulation results (Sec. IV). 
We close in Sec. V with a discussion of our findings.

\section{Model}

In Jensen's pair contact process (PCP) \cite{pcp1}, each site of the 
one-dimensional lattice ${\cal Z}$ is either vacant 
or occupied by a particle.  Each 
nearest-neighbor (NN) pair of particles has a rate $p$
of mutual annihilation, and a rate $1\!-\!p$ of attempted creation.
In a creation event involving particles at sites $i$ and $i\!+\!1$,
a new particle may appear (with equal likelihood) at site $i\!-\!1$ or
at $i\!+\!2$, provided the chosen site is vacant.
(Attempts to place a new particle at an occupied site fail.)
In an annihilation event a NN pair of particles is removed.
The PCP exhibits an active phase for $p < p_c$;
for $p \geq p_c$ the system eventually falls into an
absorbing configuration devoid of NN pairs,
but that typically contains a substantial
density, $\phi$, of isolated particles. The best
estimate for the critical parameter is $p_c \!=\! 0.077090(5)$ \cite{rdjaff}.
(Here and in what follows, numbers in parentheses denote uncertainties
in the last figure or figures.)  

To the above dynamics we now add a third process, addition of isolated
particles.  Each site is bombarded by particles at rate $h$.
(This is a fluctuating source; the mean time between successive 
addition attempts at a given site is $1/h$.)  An attempt to place a
particle at site $i$ is successful if and only if sites $i\!-\!1$, $i$, 
and $i\!+\!1$ are all vacant.  
In the absence of the source, an empty lattice is absorbing.
For any $h>0$, however, 
the insertion of isolated particles onto an initially empty
lattice corresponds to
random sequential adsorption (RSA) of dimers in one dimension.
The saturation density for this process is $(1\!-\!e^{-2})/2
= 0.432332...$ \cite{flory}.

Previous studies leave little doubt that the static critical behavior of the PCP 
(without a source) belongs to the
universality class of directed percolation.  Jensen and Dickman found that
the critical exponents $\beta$ and $\gamma$ (which govern, respectively,
the stationary mean of the order parameter, and its variance), and
$\nu_{||}$ and $\nu_{\perp}$ (which govern the divergence of the correlation time and
correlation length as one approaches the critical point),
are all consistent with DP values \cite{pcp2}.  
In addition to these static properties, the exponent
$\theta$, which governs the initial decay of the order parameter 
($\rho \propto t^{-\theta}$),
starting from a fully occupied lattice, was found consistent with DP. 
More recently, the order-parameter moment
ratios were found to be the same as those of other models belonging
to the DP universality class \cite{rdjaff}.  

Starting from a spatially homogeneous distribution of NN pairs
(for example, a completely filled lattice), the system relaxes to a
stationary state.  If, by contrast, the activity is initially localized
(e.g., a single ``seed" pair in an otherwise absorbing configuration),
we may study the spread of activity.
As noted above,
the critical exponents $\delta$ and $\eta$ characterizing spreading
vary continuously with $\phi$, and assume DP values only for
$\phi \!=\! \phi_{nat} \simeq 0.242$
\cite{rdunp,msm}.  The natural density $\phi_{nat}$ is the 
mean particle density in absorbing configurations generated by the process 
itself, at $p_c$, starting with all sites occupied.
One may, equivalently, define
$\phi_{nat}$ as the particle density in the critical {\it stationary} state, 
in the thermodynamic limit.
An environment with $\phi > \phi_{nat}$ favors spreading 
(and vice-versa), since the
higher the particle density in the environment, the more pairs will be formed per
creation event.

A kind of spreading phenomenon also arises in the stationary state
due to spontaneous fluctuations.
In the critical stationary state, we can expect to find
inactive regions of all sizes; the particle density 
in large inactive regions is $\phi_{nat}$.  When activity spreads  
into such regions, it should follow the same scaling
behavior as critical spreading with $\phi_{nat}$.  Since the exponents
governing survival and growth of activity in the stationary state 
are subject to the scaling relations 
$\delta = \beta/\nu_{||}$ and $z = 2\nu_{\perp}/\nu_{||}$,
with $\beta$, $\nu_{\perp}$ and $\nu_{||}$ 
taking DP values in the stationary state,
it follows that the spreading exponents take their usual 
DP values as well, for $\phi = \phi_{nat}$.
(In an environment with $\phi \neq \phi_{nat}$, the advance of 
activity is
no longer equivalent to that in the stationary state, 
and the spreading exponents are not constrained to take DP values.)
In this work we study {\it static} critical behavior
with $\phi \neq \phi_{nat}$ due to the action of the source.

\section{Cluster Approximations}

In this section we develop dynamic cluster approximations for the
PCP with a source of particles.  Such approximations often
yield qualitatively correct phase diagrams \cite{marro}.
For the standard PCP (no source) the two-site approximation was
presented by Carlon et al. \cite{carlon}, while the three-site approximation
was derived by Marques at al. \cite{msm}; our results for $h=0$ are
consistent with these studies.
The $n$-site approximation consists of a set of coupled differential 
equations for the probabilities $P_{\cal C}^{(n)}$ of each $n$-site 
configuration, ${\cal C}$.
(There are $2^n$ such configurations, but the number of independent
probabilities is $< 2^n$, due to normalization, and various
symmetries.)
The system is assumed homogeneous, so that the $P_{\cal C}^{(n)}$ 
are independent of position.  

Since transitions in a set of $n$ contiguous sites generally
depend on sites outside the cluster, the $n$-site probabilities
are coupled to those for $n\!+\!1$ and so on, generating an infinite
hierarchy.  The $n$-site approximation truncates this
hierarchy by approximating $m$-site probabilities (for $m>n$)
in terms of $n$-site {\it conditional} probabilitites.
For example, if $\sigma_k$ represents the state of site $k$,
then in the two-site approximation a three-site joint probability
for a sequence of nearest-neighbor sites, $k$, $l$, $m$,
is approximated so:

\begin{equation}
P(\sigma_k,\sigma_l,\sigma_m) \simeq 
P(\sigma_k|\sigma_l) P(\sigma_l,\sigma_m)    
= \frac{P(\sigma_k,\sigma_l) P(\sigma_l,\sigma_m)}{P(\sigma_l)} \;.
\label{factpa}
\end{equation}

It is convenient to denote configurations in the PCP by a string 
of 0's and 1's, the former representing vacant sites and the latter,
occupied sites.  We denote the probability of configuration 
$\sigma_1,...,\sigma_n$ by $(\sigma_1,...,\sigma_n)$, i.e., (1)
denotes the probability of a randomly chosen site being occupied,
(11) the probability of a nearest-neighbor occupied pair, and so on.

In the one-site approximation we have the transitions $(1) \to (0)$
at rate $2p(1)^2$ and $(0) \to (1)$ at rate $(1\!-\!p)(1)^2[1-(1)]
+ h[1-(1)]^3$, yielding the equation of motion:

\begin{equation}
\frac{d}{dt} (1) = -2p(1)^2 + (1\!-\!p)(0)(1)^2 + h(0)^3
\label{1site}
\end{equation}
where of course $(0) = 1 - (1)$.  
For $h=0$ the stationary active solution is
$\overline{(1)} = (1-3p)/(1-p)$ so that the critical
annihilation probability $p_c = 1/3$.  
[Aside from the active solution, we always have the absorbing state,
(1) = 0.]  For $h>0$,
there is a nonzero stationary particle density for {\it any} $p \in [0,1]$, 
i.e., the source removes the phase transition, at this level of approximation.

Next we consider the pair approximation.  There are two independent
probabilities, since, by symmetry, (10) = (01), while normalization 
implies (00) + 2(01) + (11) = 1.
There are five possible transitions of a NN pair of sites:
from (00) to (01); from (01) to (00) or (11); from (11) to
(00) or (01).  
To illustrate how rates are calculated, we 
consider the transition (00) $\to$ (01).  There are two possible
mechanisms: one involves a (11) pair just to the right of the central
pair; the other involves the action of the source, and requires a
vacant site to the right of the central pair.  In the first case, the
probability of the required configuration, (0011), is approximated
as (00)(01)(11)/[(0)(1)]; the intrinsic rate is $(1\!-\!p)/2$.  In the
second case the configuration probability is $(000) \simeq (00)^2/(0)$
in the pair approximation.  Thus the contribution to $d (01)/dt$ due to
the transition (00) $\to$ (01) is

\[
\frac{1\!-\!p}{2} \frac{(00)(01)(11)}{(0)(1)} + h \frac{(00)^2}{(0)}
\]
Note that this expression is multiplied by -2 in the equation for
$d (00)/dt$, to take the mirror-image transition [(00) $\to$ (10)]
into account.  (Recall that (10) has been eliminated by symmetry.) 

Proceeding in this manner, one readily obtains a pair of equations
for (11) and (1) = (11) + (01):

\begin{equation}
\frac{d}{dt} (11) = -p[2(11)+(1)]\frac{(11)}{(1)}
+ (1\!-\!p) [1-(11)] \frac{(11)[(1)-(11)]}{(1)(0)}
\label{2site}
\end{equation}

\begin{equation}
\frac{d}{dt} (1) = -2p(11) + (1\!-\!p)\frac{(01)(11)}{(1)} + h\frac{(00)^2}{(0)}
\label{2sitea}
\end{equation}

\noindent The stationary active solution for $h\!=\!0$ is 
$(1) \!=\! (1\!-\!5p)/(1\!-\!p)$,
$(11) = (1\!-\!3p)(1)/(1\!-\!p)$, so that $p_c = 1/5$. 
The critical annihilation rate is 2/3 for any nonzero $h$.

The three-site approximation involves five independent variables and
thirteen distinct transitions.  Integrating the coupled equations
numerically, one finds $p_c = 0.1277$ for $h=0$, and $p_c = 0.18197$
for any nonzero source.  We have analyzed cluster approximations for
up to 6 sites; the predictions for $p_c$ and for the density of isolated particles $\phi_{nat}$
at the critical point are summarized in Table I.  
The cluster approximations appear to approach the simulation values
(note the oscillatory nature of the approach to $\phi_{nat}$). 
We find in the $n=2$ approximation that for nonzero $h$, the order
parameter  $\rho \propto (p_c - p)^2$ in the neighborhood of the critical point,
that is, the mean-field exponent $\beta_{MF}=2$ in this case; for
$n > 2$ we find the usual value, $\beta_{MF} =1$.

For $h=0$ we can obtain a rough estimate of $p_c$ by extrapolating the cluster
results, i.e., via a linear fit to $p_c (n)$ plotted versus $1/n$.  Using the data for
$n = 3 - 6$ yields $p_c = 0.082(7)$; using only the $n=4 -6$ data, we find
$p_c = 0.07(1)$, consistent with the simulation result.
Unfortunately the behavior of these low-order approximations is not sufficiently
regular to allow an extrapolation to $n \to \infty$, when $h>0$.

Given a sequence of cluster approximations, and knowing the value of $p_c$,
one may apply Suzuki's {\it coherent anomaly} analysis to extract certain
critical exponents \cite{cam}.  In the present case, however, our estimates
for $\beta$ appear to converge slowly; we find
$\beta \simeq 0.29$ for $h=0$ (from a quadratic fit to the order parameter
at the critical point),
and $\simeq 0.28$ for $h=0.5$.  Data for
larger clusters will be needed in order to derive precise predictions for
critical exponents.

In Fig. 1 we compare the $n$-site approximation predictions against simulation
results for the phase boundary, $p_c(h)$.  The cluster approximations appear
to approach the simulation curve in a qualitative fashion.
On the other hand, all the approximations ($n \leq 6$) predict a jump in 
$p_c$ at $h\!=\!0$, while simulations show $p_c$ to be continuous, though singular,
at this point.  Thus it appears that detailed features of the phase diagram
are beyond the small-cell approximations developed here.

\section{Simulations}

\subsection{Method}

We first define the simulation algorithm for the PCP in 
the absence of a source
($h\!=\!0$).  We consider a ring of $L$ sites.
Since all events depend upon the presence of a NN pair, we
maintain a list of such pairs.  An event consists in (1) choosing a process
(annihilation with probability $p$, creation with probability $1\!-\!p$);
(2) choosing a NN pair at random from the list; (3) deleting the two
particles in the pair, in case of annihilation, or, in case of creation,
adding a new particle adjacent to the pair, if possible.  More specifically,
in the case of creation, step (3) consists in choosing (with equal
probabilitites) the site immediately to the left, or to the right, of the
NN pair, and inserting a particle at this site if it is vacant.
The time increment associated with each event (creation or annihilation)
is $\Delta t = 1/N_p$ where $N_p$ is the number of NN pairs just prior
to the event.  (If $N_p=0$ the system has fallen into an absorbing
configuration and the trial ends.)
In this way each NN pair undergoes, on average, one
event (creation or annihilation) per unit time.  The list of NN pairs must 
naturally be updated following each annihilation or sucessful creation.

When $h>0$, each creation or annihilation is followed by a
certain number $N_h$ of source events.  Suppose
a creation or annihilation has just occured, and that there are
now $N_p$ pairs.  Then the next creation or annihilation will 
occur after a time interval of $1/N_p$; during this interval, the
expected number of source events is $Lh\Delta t = Lh/N_p$, since $h$
is the rate of insertion attempts per site and unit time.  Thus after
each creation or annihilation we perform $N_h$ insertion attempts, 
where $N_h$ is a Poissonian random variable
with mean $Lh/N_p$.
An insertion attempt consists in choosing a site $i$ at random (each site
has probability $1/L$; no lists are used in this process), and
inserting a particle at there if and only if sites $i-1$, $i$, and $i+1$ are
all vacant.

\subsection{Results}

We used the simulation algorithm described above to determine
$p_c (h)$ at a number of different $h$ values.  In addition, detailed
studies were performed at $h=0$, 0.1, and 10 in order to obtain precise
values for critical exponents and other properties.  All simulations
start from a fully occupied lattice.  The order parameter $\rho = N_p/L$
(i.e., the density of nearest-neighbor pairs), 
the survival probability, $P(t)$ (that is,
to have $N_p > 0$), and the particle density, $\psi$, are
monitored during the evolution, and, in particular, in the stationary state.
The properties of principal interest as regards critical
behavior are the stationary
order parameter $\overline {\rho}$ (the overline denotes a stationary
mean), the moment ratio 
$m \equiv \overline {\rho^2}/(\overline {\rho})^2$, and the lifetime
$\tau$, obtained by fitting an exponential 
to the survival probability: $P(t) \sim e^{-t/\tau}$. 
Uncertainty estimates for $\overline {\rho}$, $\overline{\psi}$, $m$ and
$\tau$ are obtained from the standard deviation evaluated over
a set of 3-5 independent runs.

In order to fix the critical point, $p_c$, for a given value of $h$,
we employ three criteria.  The first is the finite-size scaling behavior
of the order parameter: one expects a simple power-law dependence 
at the critical point, 

\begin{equation}
\overline {\rho} (p_c,L) \sim L^{-\beta/\nu_{\perp}} \;,
\label{fssrho}
\end{equation}
as has in fact been verified for many absorbing-state phase 
transitions \cite{marro,hinr00}.  For noncritical $p$-values,
the dependence of $\overline {\rho} (p_c,L) $ upon $L$ shows
deviations from a power law, typically manifested as curvature on
a plot of $\overline {\rho} (p_c,L)$ versus $L$ on log scales. 
The same considerations apply to the survival time, whose
critical finite-size scaling form is

\begin{equation}
\tau (p_c,L) \sim L^{\nu_{||}/\nu_{\perp}} \;.
\label{fsstau}
\end{equation}
Our second criterion for locating $p_c$ is thus power-law
dependence of $\tau$ on the system size.

A third criterion is based on the independence of moment ratios
such as $m$ at the critical point.  This property, which has long been
used in studies of equilibrium critical phenomena \cite{binder},
was more recently verified for absorbing-state phase transitions
in the contact process and PCP \cite{rdjaff,jaffrd}.  We
determine the value $p^*(L_1,L_2)$ at which the moment ratios
$m(p,L_1)$ and $m(p,L_2)$ take the same value, for pairs of
system sizes $L_1$ and $L_2$.  Extrapolating $p^*$ to $L \to \infty$
yields an estimate for $p_c$.

We studied system sizes $L=50$, 100, 200, 500, and 1000.  
A study at a particular value of $h$ begins with a quick survey of
small systems, to obtain a preliminary estimate of $p_c$.  Then, for
each system size, we perform high-statistics studies at three to five
$p$-values, obtaining the properties of interest to high precision.
Values of $\overline {\rho}$, $m$, $\tau$, and $\overline{\psi}$ at
intermediate $p$-values are obtained via interpolation of the
high-precision data using polynomial (typically cubic) least-squares
fits.  

We tested this procedure on the standard PCP ($h\!=\!0$).  For $L\!=\!50$
we ran a total of $6 \times 10^5$ trials at $p$-values of
0.0769, 0.0770, 0.0771, and 0.0772.  For this small system the
stationary state is well sampled in trials that extend to a maximum time
of 500.   For $L\!=\!1000$, we performed a total
of $3 \times 10^4$ trials (extending to a maximum time of $10^5$)
at $p=0.0770$, 0.0771, and 0.0772.  Sample sizes and maximum times
for intermediate system sizes fall between the values quoted for
$L\!=\!50$ and 1000;  similar parameters were used in the 
high-precision studies at $h\!=\!0.1$ and 10.
Figure 2 illustrates the results of the fitting procedure for
$\rho$; it is clear that the simulation data are well represented
by a smooth function.  In Fig. 3 we show all of the data for $m$
(for $h\!=\!0$) together with the associated polynomial fits.

For $h\!=\!0$, the three criteria mentioned above yield
$p_c = 0.077091(5)$, in excellent agreement with previous studies
\cite{rdjaff,rdunp,pgunp}.  The uncertainty in $p_c$ reflects
scatter in the last three moment-crossing points, and in the
p-value yielding power laws for $\overline{\rho}$ and for
$\tau$.  Extrapolating the values of $m(p_c,L)$
to $L \to \infty$, we find $m=1.1740(5)$. 
(Here the uncertainty includes three sources: that in the original
data, the uncertainty of the extrapolation at $p_c$, and the
uncertainty in $p_c$ itself.  The same applies to the exponent
values discussed below.)
Our result is once again in excellent
agreement with that of Ref. \cite{rdjaff},
$m_c=1.1735(5)$ for transitions in the DP universality class
in 1+1 dimension.  

From the scaling of $\overline{\rho}$ at the
critical point we find $\beta/\nu_\perp = 0.2522(5)$ (see Fig.~4); this
exponent ratio takes the value 0.2521 for the DP class \cite{jensen96}.
The survival time data (Fig. 5) yield $\nu_{||}/\nu_\perp = 1.577(4)$, while
the DP value is 1.5807 \cite{jensen96}.  
As noted in the Introduction,
the order parameter is expected to follow a power law, 
$\rho \sim t^{-\theta}$,
during the approach to the stationary state.
A study at $p=0.077091$ in a
system of 1000 sites yielded $\theta = 0.1596(2)$ (see Fig.~6),
while the expected DP value is $\theta = \beta/\nu_{||} = 0.15947(3)$.
Finally, a study of 
$\overline {\rho}$ in the supercritical regime ($p = 0.0722$ - 0.076,
for $L=1000$) yielded $\beta = 0.276(3)$
[via the usual relation, $\overline{\rho} \sim (p_c-p)^\beta$].
The accepted value is $\beta = 0.27649(4)$ \cite{jensen96}.
We note that these exponent values are the highest precision yet
reported for the PCP, and that they place the model unambiguously
in the DP universality class.  

While the results for $\beta/\nu_\perp$ and $\beta$ furnish the
estimate $\nu_\perp = 1.092(7)$, it is desirable, for reasons 
explained below, to have an independent estimate of $\nu_\perp$.
Since $\Delta \!=\! p_c-p$ enters all finite-size scaling forms in the
combination $\Delta L^{1/\nu_\perp}$, we can obtain such an estimate
either from a data-collapse analysis, or by studying the finite-size
dependence of a derivative such as $dm/dp$ at $p_c$; the latter must diverge
as  $L^{1/\nu_\perp}$.  We find that this derivative does follow a
power law; a fit to the data yields $\nu_\perp = 1.086(22)$, where the
relatively large error reflects the uncertainty associated
with numerical evaluation of a derivative.  (Both our estimates are consistent with the
DP value, $\nu_\perp = 1.0968$.)

Our data for the particle density $\overline{\psi} (p_c,L)$ fall on a straight line when 
plotted versus $L^{-\beta/\nu_\perp}$, and extrapolate to 0.241(1)
for $L \!\to\! \infty$ (Fig.~7).  In the active state, $\overline{\psi}$ 
represents the {\it total}
particle density (isolated particles as well as those belonging to
pairs), but when $L \!\to\! \infty$ at $p\!=\!p_c$ there are no pairs; in this
limit $\overline{\psi}$ corresponds to
the natural density, $\phi_{nat}$,
of isolated particles.  Our result agrees with previous studies,
based directly on absorbing configurations \cite{rdunp,msm}.

The procedure described above, which furnishes results of unprecedented
precision for the standard PCP, was repeated 
for $h\!=\!0.1$ and 10.  The resulting
critical parameters are compared with the $h\!=\!0$ case in Table II.
We remark that the dependence of $\overline{\rho}$, $\tau$, and $m$
on $L$, at the critical point, is qualitatively the same as for $h\!=\!0$,
and that the associated critical exponents,
$\beta/\nu_\perp$, $\nu_{||}/\nu_\perp$, and $\theta$, and 
moment ratio $m$, are the same, to
within uncertainty, as for $h\!=\!0$.  
This is particularly clear from Fig. 4, which compares the dependence
of $\overline{\rho}$ on $L$ at the critical point for $h=0$,
0.1, and 10, and Fig. 5, a similar plot of the lifetime $\tau$.  
The same power laws are seen, regardless of the value of $h$,
the sole difference between zero and nonzero $h$ being a slight
change in amplitude.  The data for $h=0.1$ and $h=10$ are virtually
identical.

The decay of $\rho$ at the critical point again appears to follow
the same power law, regardless of whether or not the source is present.
Fig. 6  shows that the main effect of the source is a small decrease
in the amplitude.  Analysis of the local slopes of the graphs in Fig. 6
leads to $\theta = 0.1603(5)$ in the presence of the
source, in very good agreement with the value 
$\theta = 0.1596(4)$ found for $h\!=\!0$.
The rather imprecise estimates
for $\nu_\perp$, obtained directly from the scaling of $dm/dp$,
likewise agree to within uncertainty, independent of $h$.

There are two principal differences between the scaling properties
with and without the source.  First, Table II shows that the
limiting ($L \!\to\! \infty$) value of the particle density 
$\overline{\psi}$ is
much larger when $h>0$.  (The data for $\overline{\psi}$ 
at the critical point, 
and the extrapolation to $\psi_\infty$, are depicted in Fig. 7.)
In fact, there is good reason to expect
$\psi_\infty$ to jump from $\phi_{nat}$ to a higher value as soon
as $h>0$ (just as in the cluster approximations of the previous section),
since, no matter how weak, the source will eventually fill in all
available sites (as in RSA), in the absence of activity.  With increasing $h$,
$\psi_\infty$ approaches the RSA value of 0.4323.  

In addition to the expected difference in limiting 
particle densitites,
the source appears to induce a rather subtle change in the critical
exponent $\beta$ (see Fig.~8).  The value for 
$h>0$ is about 4\% greater than for $h\!=\!0$; since only about 2\% of this
difference can be attributed to uncertainty, this seems to represent a
significant, albeit small, change in the exponent.  The interpretion
this result, however, is not straightforward, since all of
the other exponents (or ratios) studied appear to be insensitive to
the source.  In particular, the values for 
$\beta/\nu_\perp$ are constant to within 1.2\%, while our result
for $\nu_\perp$ (from the scaling of $dm/dp$) is constant to within 
0.6\%, implying constancy of $\beta$ to better than 2\%.  Thus it
is difficult to accept the apparent change in $\beta$ at face value;
a better understanding will require either a theory of the
effect of the source, or numerical results for larger systems.  It is
worth noting that with the exception of $\beta$, all exponents have
been obtained via finite-size scaling at the critical point.  Thus our
results are consistent with the possibility that the source modifies 
the apparent scaling of the order parameter near, but not at, the 
critical point, perhaps due to additional corrections to scaling.  
We cannot
rule out a true (as opposed to merely apparent) change in the exponent
$\beta$ with the data in hand, but, given our other results, such a
modification would entail a very surprising violation of the usual
connection between scaling in the supercritical regime and
finite-size scaling at the critical point.

While the effect of the source on critical exponents, if any, is
quite subtle, its effect on the position of the critical point, $p_c$,
is dramatic.  Fig. 9 shows the phase boundary in the $h$-$p$ plane,
as determined in simulations.  Evidently $p_c$ grows in a singular
manner as $h$ is increased from zero.  Although there is no jump in
$p_c$, (contrary to the predictions of the cluster approximations
described in the preceding
section), all evidence points to the derivative $dp/dh$ being infinite
at $h=0$.  For larger $h$, $p_c$ saturates; plotting $p_c$
versus $1/h$ leads to the estimate $p_c = 0.0966(1)$ in the $h \!\to\! \infty$
limit.  The nature of the singularity at $h\!=\!0$ is not fully clear:
a double logarithmic plot of $\Delta p_c = p_c(h) - p_c(0)$ versus $h$
does not yield a simple power law, perhaps due to saturation-induced
curvature, but suggests an exponent of $\approx 1/2$ as $h \to 0$ (see Fig. 9, inset).
As can be seen from Fig. 1, it is plausible that $\Delta p$ is
a smooth function of $h^{1/2}$.  Elucidating the precise nature
of the singularity is a central issue for future study.

\section{Discussion}

Motivated by the question of universality at an absorbing-state phase 
transition
in the presence of a source coupling
to a non-ordering field, we investigated the phase diagram and
critical behavior of the PCP with a source of isolated particles.
We studied the model via cluster approximations and extensive 
simulations.
$n$-site cluster approximations (for $n \leq 6$) yield
predictions for the phase boundary $p_c(h)$, and
the isolated particle density $\phi$, that appear to approach
the simulation results.
 All of the cluster
approximations studied here predict of a
discontinuity in $p_c$ at $h=0$; simulations show $p_c(h)$
to be a continuous function, albeit singlar at $h=0$.

The central conclusion from extensive simulations performed for source 
strengths $h=0$, 0.1, and 10 is that the presence of the source, and the
associated change in the background density of isolated particles,
has no detectable effect on scaling at the critical point.  The present
level of precision permits us to state that the values of 
$\nu_\perp$, $\beta/\nu_\perp$, $\nu_{||}/\nu_\perp$, $\theta$,
and the moment ratio $m = \overline {\rho^2}/(\overline {\rho})^2$,
are constant to within 1\% as we vary the intensity of the source.
Our results for these parameters are fully consistent with known values
for the DP universality class.
We have noted, on the other hand, that in the presence of the source,
the critical exponent $\beta$ 
appears to be about 4\% greater than the DP value. 
This is in conflict with the balance of our results
which indicate constancy of the critical exponents.  
A resolution must await development of a theoretical understanding of 
scaling in the presence of the source, and/or more extensive simulations.

The apparent insensitivity of the critical behavior to a change in the stationary
density places the PCP in the same category as the threshold transfer process
(TTP) \cite{mendes,ttpnote}.  Whereas in the TTP the nonordering field can relax in the
absence of activity, while in the PCP it cannot, the two models are similar in that
the nonordering field exhibits only short-range correlations \cite{pcp2,snr}.  
An interesting open question is whether perturbations in a nonordering field 
that lead to long-range
correlations can affect critical behavior.

Our simulations reveal that the phase boundary, $p_c(h)$, is singular 
at $h=0$.  The available data are consistent with an asymptotic power 
law $\Delta p_c \sim h^{\alpha}$ with $\alpha \approx 1/2$, but further
studies will be needed to characterize the singularity precisely.
A scaling or renormalization-group analysis of the PCP with an isolated
particle source would clearly be desirable, in order to understand the
form of the phase diagram, and the scaling of the order parameter.
\vspace{2em}

\noindent {\bf Acknowledgments}
\vspace{1em}

We are grateful to Miguel Angel Mu\~noz for valuable discussions, and to
Peter Grassberger and Maria Augusta Santos, for communicating their
results prior to publication.
This work was supported by CNPq and FAPEMIG.  G.O. acknowledges support
from Hungarian Research Fund OTKA (T-25286), and Bolyai (BO/00142/99).
\vspace{1.5em}

\noindent 
$^1${\small electronic address: dickman@fisica.ufmg.br } \\
$^2${\small electronic address: odor@mfa.kfki.hu } \\

\newpage

\begin{table}
\caption{\sf $n$-site approximation results}
\begin{center}
\begin{tabular}{|c|c|c|c|c|} 
$n$ &$p_c (h\!=\!0)$ & $p_c (h\!=\!0.1)$ & $\phi (h\!=\!0)$&$\phi (h\!=\!0.1)$ \\
\hline\hline
2    &       0.2000     &         0.6667       &       0              & 0.5   \\
3    &       0.1277     &         0.1820       &       0.23         & 0.461 \\
4    &       0.1185     &         0.15023     &       0.2166      & 0.5    \\
5    &       0.114       &         0.14405     &       0.18          & 0.46  \\
6    &       0.103       &         0.117        &       0.2234       & 0.438  \\
\hline\hline
{\small SIM} & 0.077091(5)  & 0.086272(15) & 0.241(1)  & 0.421(1) \\
\end{tabular}
\end{center}
\label{nsite}
\end{table}

\begin{table}
\caption{\sf Critical parameters of DP and the PCP.
DP exponents from Ref. [47], $m$ from Ref. [32].
Numbers in parentheses denote uncertainties in the last figure(s).}
\begin{center}
\begin{tabular}{|c|c|c|c|c|c|c|c|c|} 
$h$ &$p_c$ &$\beta/\nu_\perp$&$\nu_{||}/\nu_\perp$&$\beta$&$\nu_\perp$&$\theta$ & $m$ &$\psi_\infty$ \\
\hline\hline
DP  &      & 0.25208(4) & 1.5807 & 0.27649 & 1.09684(1) & 0.15947(3) & 1.1735(5) &      \\
\hline
0   & 0.077091(5) & 0.2523(3) & 1.577(4) & 0.276(3)& 1.086(22) & 0.1596(4) & 1.1740(5) & 0.241(1)  \\
0.1 & 0.086272(15)& 0.2540(15)& 1.571(9) & 0.287(3)  & 1.083(20) & 0.1602(4) & 1.176(1)& 0.421(1) \\
10  & 0.097850(10)& 0.2554(10)& 1.574(10)& 0.287(3)  & 1.080(10) & 0.1604(4) & 1.175(1)& 0.433(1)  \\
\end{tabular}
\end{center}
\label{pcprat}
\end{table}

\newpage

\noindent {\bf Figure Captions}
\vspace{1em}

\noindent Fig. 1.  Cluster approximation predictions (for $n\!=\!4$, 5, and 6,
as indicated) and simulation results (circles) for the phase boundary.
\vspace{1em}

\noindent Fig. 2.  Simulation data for the stationary order
parameter value, $\overline{\rho}$, for $L=50$ and $h=0$.
The solid line is a cubic fit.
\vspace{1em}

\noindent Fig. 3.  Simulation data and polynomial
fits for the moment ratio $m$, for $h=0$.  System sizes
$L=50$, 100, 200, 500 and 1000 in order of increasing slope.
Inset: detail of crossing region for $L=100$,...1000.
\vspace{1em}

\noindent Fig. 4.  Stationary density versus system size at the
critical point for $h=0$ ($\Box$), $h=0.1$ (+), and $h=10$ ($\diamond$).
Error bars are smaller than the symbols.
\vspace{1em}

\noindent Fig. 5.  Lifetime versus system size at the critical point.
Symbols as in Fig. 4.
\vspace{1em}

\noindent Fig. 6.  Decay of the order parameter at $p_c$.  Upper curve:
$h=0$; dotted line $h=0.1$; lowermost curves: $h=1$ and $h=10$, 
which cannot be distinguished on this scale
\vspace{1em}

\noindent Fig. 7.  Stationary particle density versus
$L^{-\beta/\nu_\perp}$, illustrating linear extrapolation to
$\psi_\infty$.  Symbols as in Fig. 4.
\vspace{1em}

\noindent Fig. 8.  Stationary pair density versus
$\Delta = p_c-p$, for $h=0$, 0.1, and 10; $L\!=\!1000$.  Symbols as in Fig. 4.
Straight lines represnet least-squares fits to the linear portion of the
data for $h=0$ and $h=10$; the slopes are 0.276 and 0.287, respectively.
\vspace{1em}

\noindent Fig. 9.  Critical annihlation rate $p_c$ versus source
strength $h$  (the dashed line is simply a guide to the eye).  
The inset shows the critical point shift $\Delta p_c$ versus
$h$ on log scales; the straight line has a slope of 1/2.
\vspace{1em}


\newpage
\begin{thebibliography}{99}

\bibitem{privbook} 
     R. Dickman, in 
     {\it Nonequilibrium Statistical Mechanics in One Dimension}, 
     edited by V. Privman, (Cambridge University Press, Cambridge, 1996).

\bibitem{marro} 
     J. Marro and R. Dickman,
     {\em Nonequilibrium Phase Transitions in Lattice Models} 
     (Cambridge University Press, Cambridge, 1999).

\bibitem{hinr00}
     H. Hinrichsen,
     Adv. Phys. {\bf 49}, 815 (2000).

\bibitem{harris} 
     T. E. Harris, 
     Ann. Prob. {\bf 2}, 969 (1974).

\bibitem{zgb} 
     R. M. Ziff, E. Gulari, and Y. Barshad, 
     Phys. Rev. Lett. {\bf 56}, 2553 (1986).
     
\bibitem{evans} 
     J. W. Evans, 
     Langmuir {\bf 7}, 2514 (1991).

\bibitem{alon} 
      M. Alon, M. R. Evans, H. Hinrichsen, and D. Mukamel,  
      Phys. Rev. Lett. {\bf 76}, 2746 (1996).

\bibitem{gz} 
     P. Grassberger and Y. C. Zhang, 
     Physica A {\bf 224}, 169 (1996).

\bibitem{maslov} 
        S. Maslov, M. Paczuski, and P. Bak, 
        Europhys. Lett. {\bf 27}, 97 (1994).

\bibitem{dvz} 
     R. Dickman, A. Vespignani, and S. Zapperi,
     Phys. Rev. E {\bf 57}, 5095 (1998).

\bibitem{vdmz} 
     A. Vespignani, R. Dickman, M. A. Mu\~noz, and S. Zapperi,
     Phys. Rev. Lett. {\bf 81}, 5676 (1998).

\bibitem{pomeau}
      Y. Pomeau, 
      Physica D {\bf 23}, 3 (1986).

\bibitem{chate}
      H. Chat\'e and P. Manneville,
      Phys. Rev. Lett. {\bf 58}, 112 (1986).

\bibitem{bohr}
     T. Bohr, M. van Hecke, R. Mikkelsen, and M. Ipsen,
     Phys. Rev. Lett. {\bf 86}, 5482 (2001).

\bibitem{janssen}
     H. K. Janssen,
     Z. Phys. B {\bf 42}, 151 (1981).

\bibitem{gr1}
     P. Grassberger,
     Z. Phys. B {\bf 47}, 365 (1982).

\bibitem{glb}
     G. Grinstein, D.-W. Lai, and D. Browne,
     Phys. Rev. A {\bf 40}, 4820 (1989).

\bibitem{gkt}
     P. Grassberger, F. Krause, and T. von der Twer,
     J. Phys. A {\bf 17}, L105 (1984); P. Grassberger, 
     J. Phys. A {\bf 22}, L1103 (1989).

\bibitem{baw1}
     H. Takayasu and A. Yu. Tretyakov,
     Phys. Rev. Lett. {\bf 68}, 3060 (1992).

\bibitem{iwanbaw}
     I. Jensen,     
     Phys. Rev. E {\bf 50}, 3623 (1994).
     
\bibitem{cardybaw}
     J. L. Cardy and U. T\"auber,
     Phys. Rev. Lett. {\bf 77}, 4780 (1996).

\bibitem{benav}
     J. K\"ohler and D. ben-Avraham,
     J. Phys. A {\bf 24}, L621 (1991).

\bibitem{albdd}
     E. V. Albano,
     J. Phys. A {\bf 25}, 2557 (1992);
     J. Phys. A {\bf 27}, 431 (1994).

\bibitem{ijnoco}
     I. Jensen,
     J. Phys. A {\bf 27}, L61 (1994).

\bibitem{pcp1} 
     I. Jensen, 
     Phys. Rev. Lett. {\bf 70}, 1465 (1993).

\bibitem{pcp2} 
     I. Jensen and R. Dickman, 
     Phys. Rev. E {\bf 48}, 1710 (1993).

\bibitem{mendes}
     J. F. F. Mendes, R. Dickman, M. Henkel, and M. C. Marqu\'es,
     J. Phys. A {\bf 27}, 3019 (1994).

\bibitem{inas}
     M. A. Mu\~noz, G. Grinstein, R. Dickman, and R. Livi, 
     Phys. Rev. Lett. {\bf 76}, 451 (1996);
     Physica D {\bf 103}, 485 (1997). 

\bibitem{snr}
     R. Dickman,
     Phys. Rev. E {\bf 53}, 2223 (1996).

\bibitem{mgd}
     M. A. Mu\~noz, G. Grinstein, and R. Dickman, 
     J. Stat. Phys. {\bf 91}, 541 (1998).

\bibitem{gcr} 
     P. Grassberger, H. Chat\'e, and G. Rousseau,
     Phys. Rev. E {\bf 55}, 2488 (1997).

\bibitem{rdjaff} 
      R. Dickman and J. Kamphorst Leal da Silva,
      Phys. Rev. E {\bf 58}, 4266 (1998).

\bibitem{lopez} 
      C. L\'opez and M. A. Mu\~{n}oz, 
      Phys. Rev. E {\bf 56}, 4864 (1997).

\bibitem{odor88}
     G. \'Odor, J. F. Mendes, M. A. Santos, and M. C. Marques,
     {\it Phys. Rev. E} {\bf 58}:7020 (1998).

\bibitem{lipowski}
     A. Lipowski and M. $\not{\mbox{L}}$opata,
     Phys. Rev. E {\bf 60}, 1516 (1999).

\bibitem{rwmpr}
        R. Dickman and D. ben-Avraham,
        Phys. Rev. E {\bf 64}, 020102(R), (2001).

\bibitem{odor00}
        G.\'Odor and N. Menyh\'ard,
        Phys. Rev. E {\bf 61}, 6404, (2000).

\bibitem{ttpnote}
        In the {\it threshold transfer process} studied in Ref. \cite{mendes},
        the stationary value of the nonordering field can be varied by means
        of a parameter, but it relaxes to stationary value even in the absence
        of activity.  This variation causes a shift in $p_c$ but does not alter
        the critical exponents.

\bibitem{flory}
         P. J. Flory, 
         J. Am. Chem. Soc. {\bf 61}, 1518 (1939).

\bibitem{rdunp}
          R. Dickman, e-print: cond-mat/9909347.

\bibitem{msm}
          M. C. Marques, M. A. Santos, and J. F. F. Mendes,
          preprint.

\bibitem{carlon}
          E. Carlon, M. Henkel, and U. Schollw\"ock,
          Phys. Rev. E {\bf 63}, 036101 (2001).

\bibitem{cam}
          M. Suzuki,
          J. Phys. Soc. Jpn. {\bf 55}, 4205 (1986).

\bibitem{binder} 
          K. Binder,
          Phys. Rev. Lett. {\bf 47}, 693 (1981);
          Z. Phys. B {\bf 43}, 119 (1981).

\bibitem{jaffrd}
      J. Kamphorst Leal da Silva and R. Dickman
      Phys. Rev. E {\bf 60}, 5126 (1999).

\bibitem{pgunp}
          P. Grassberger, private communication.

\bibitem{jensen96}
          I. Jensen,
          J. Phys. A {\bf 29}, 7013 (1996).

\end{thebibliography}
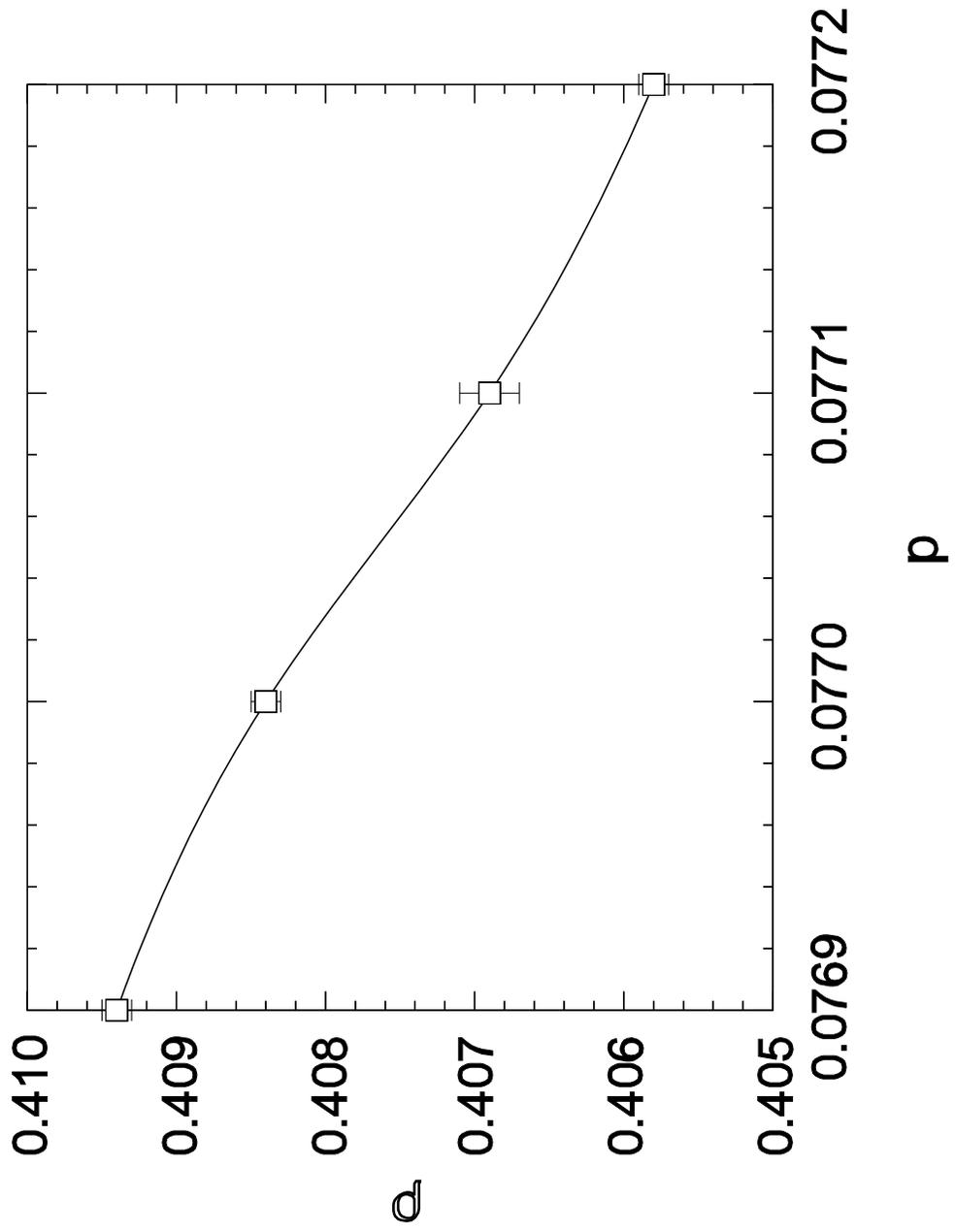
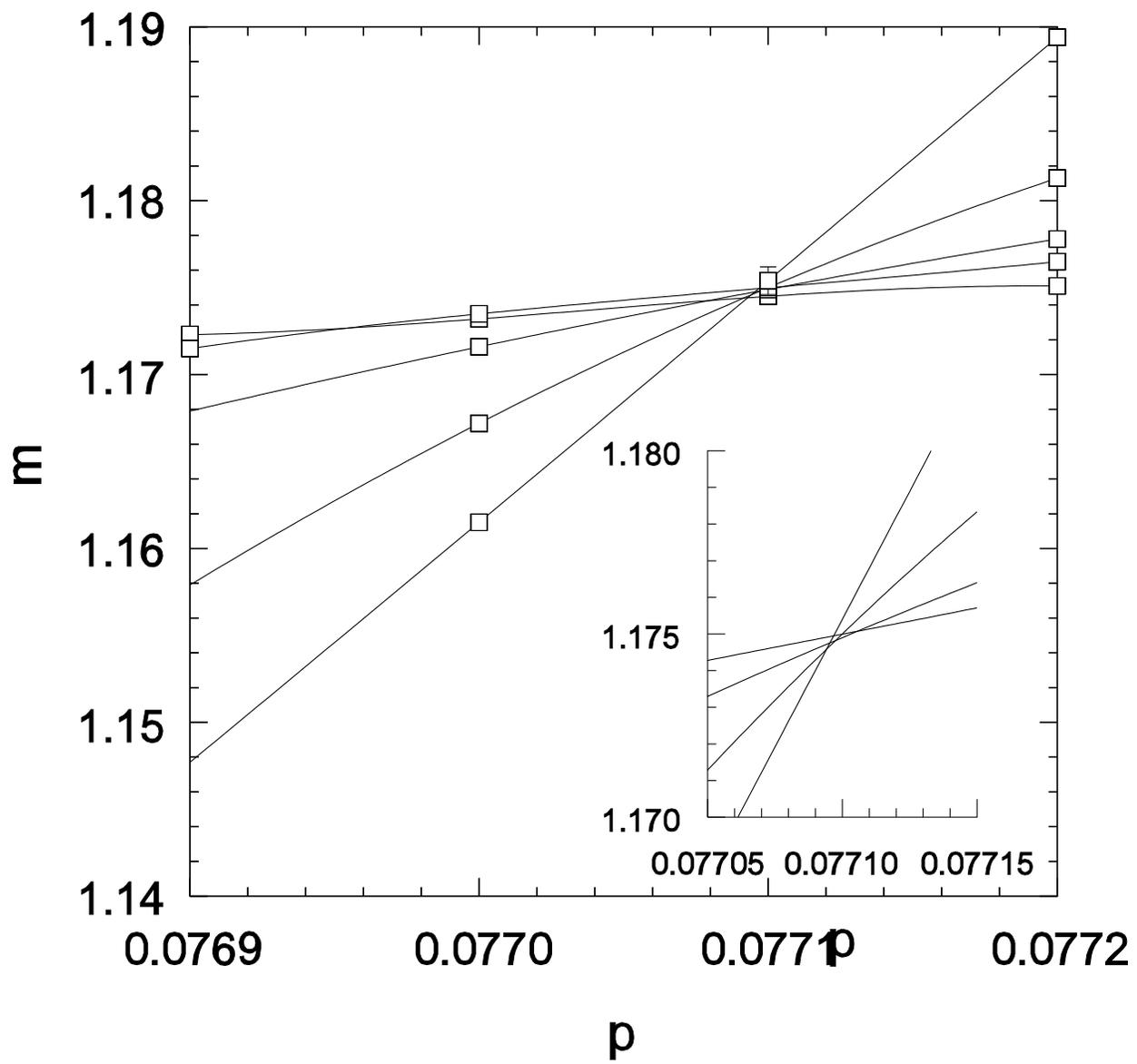
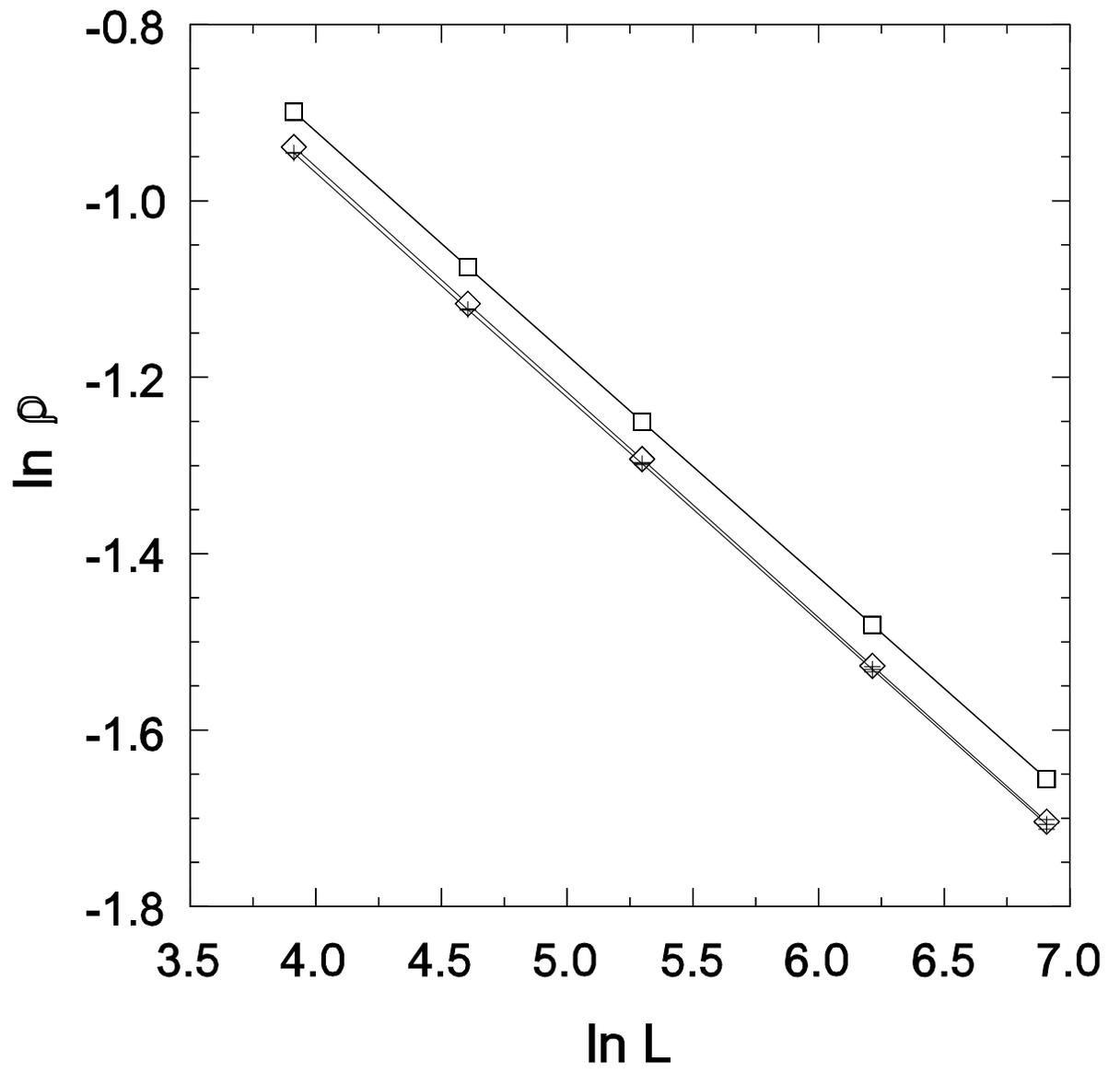
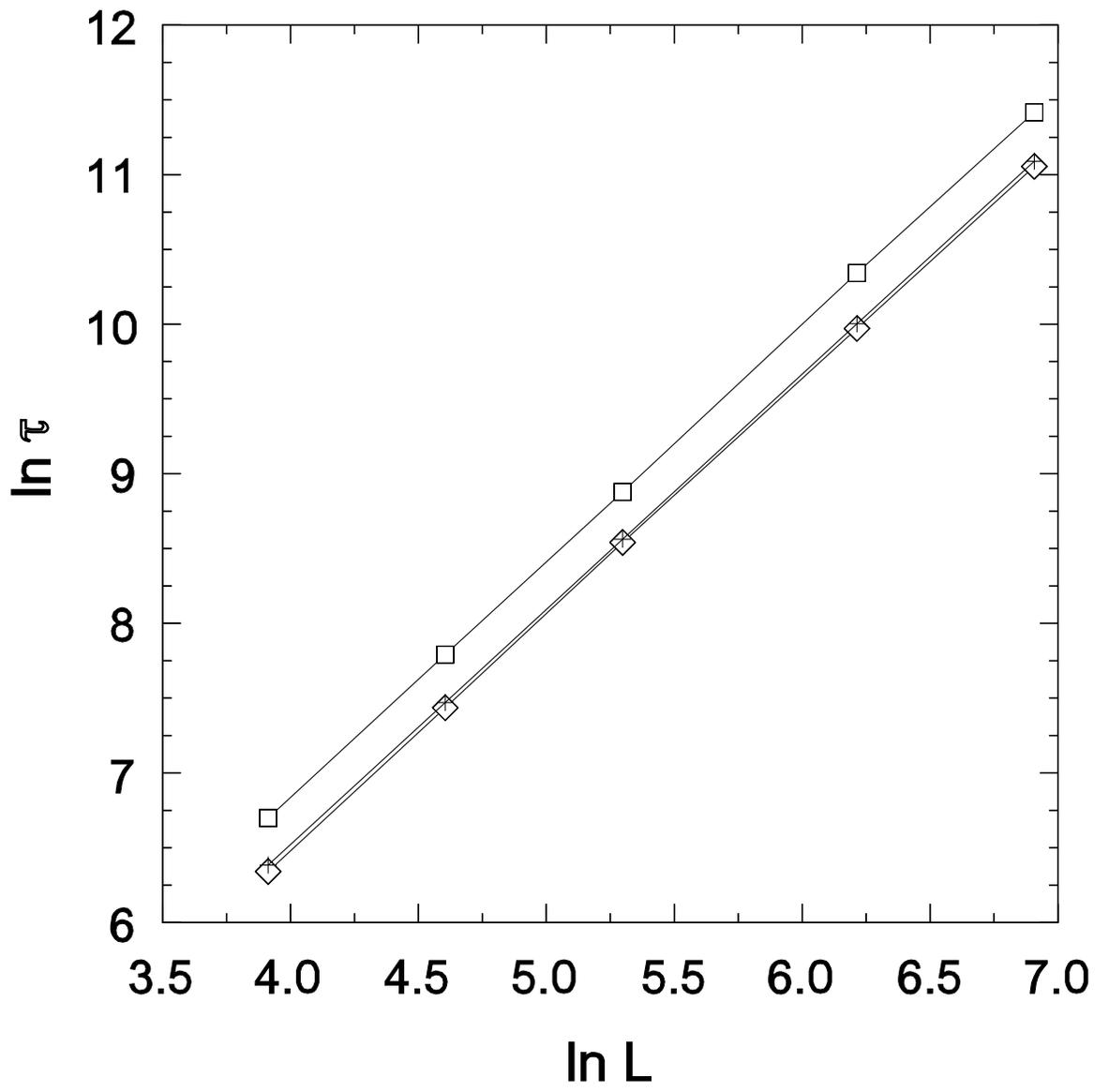
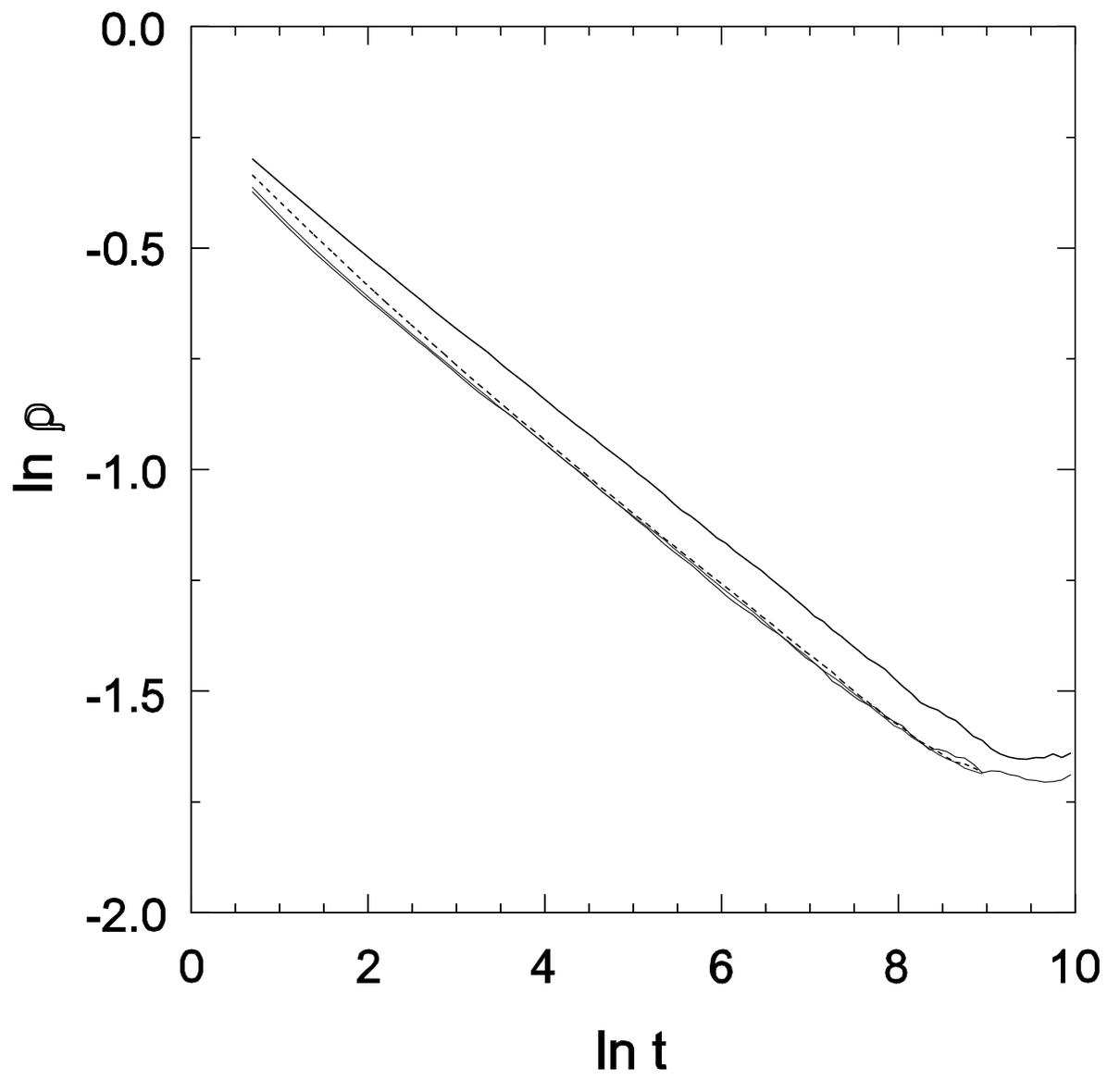
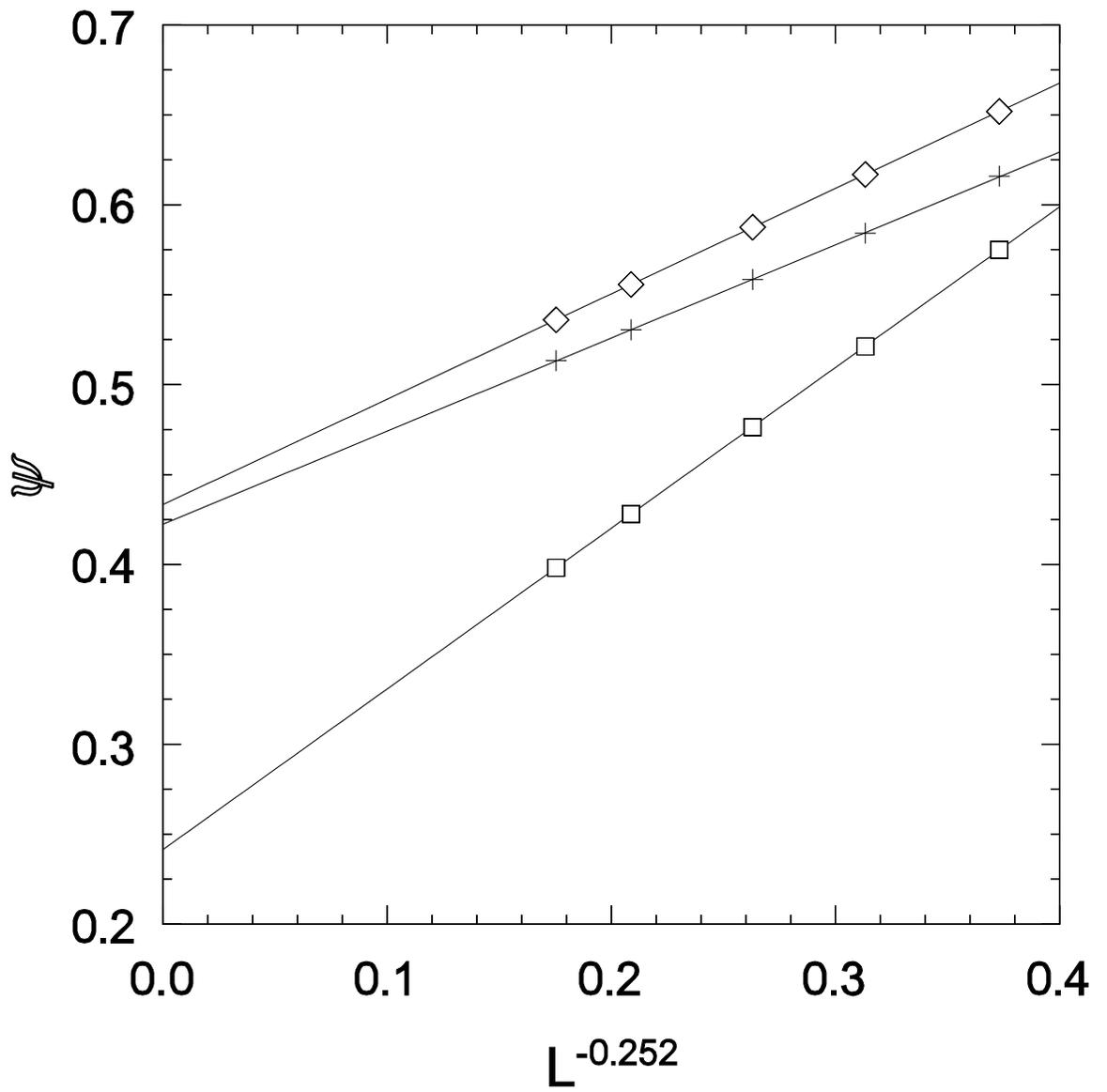
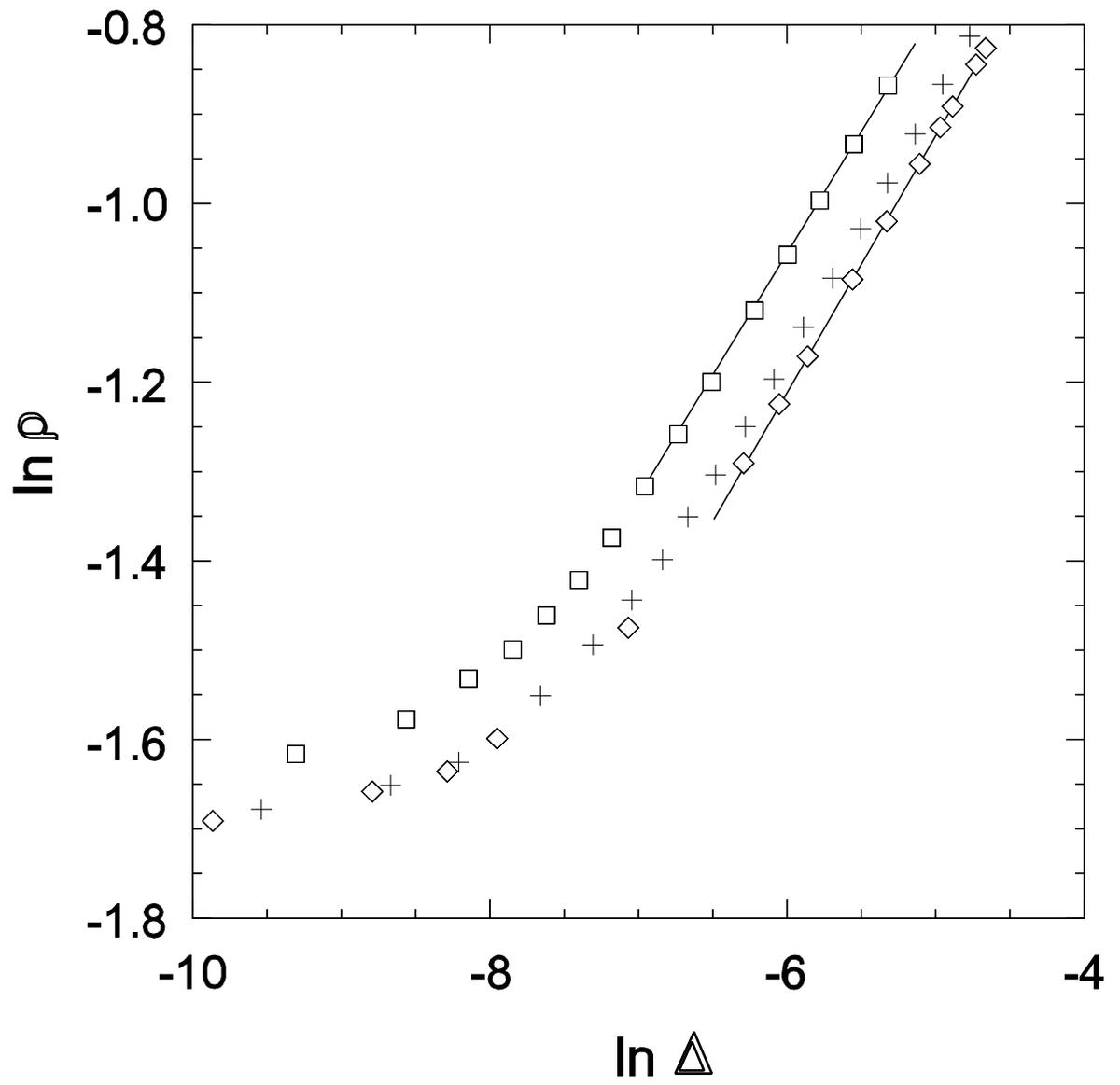
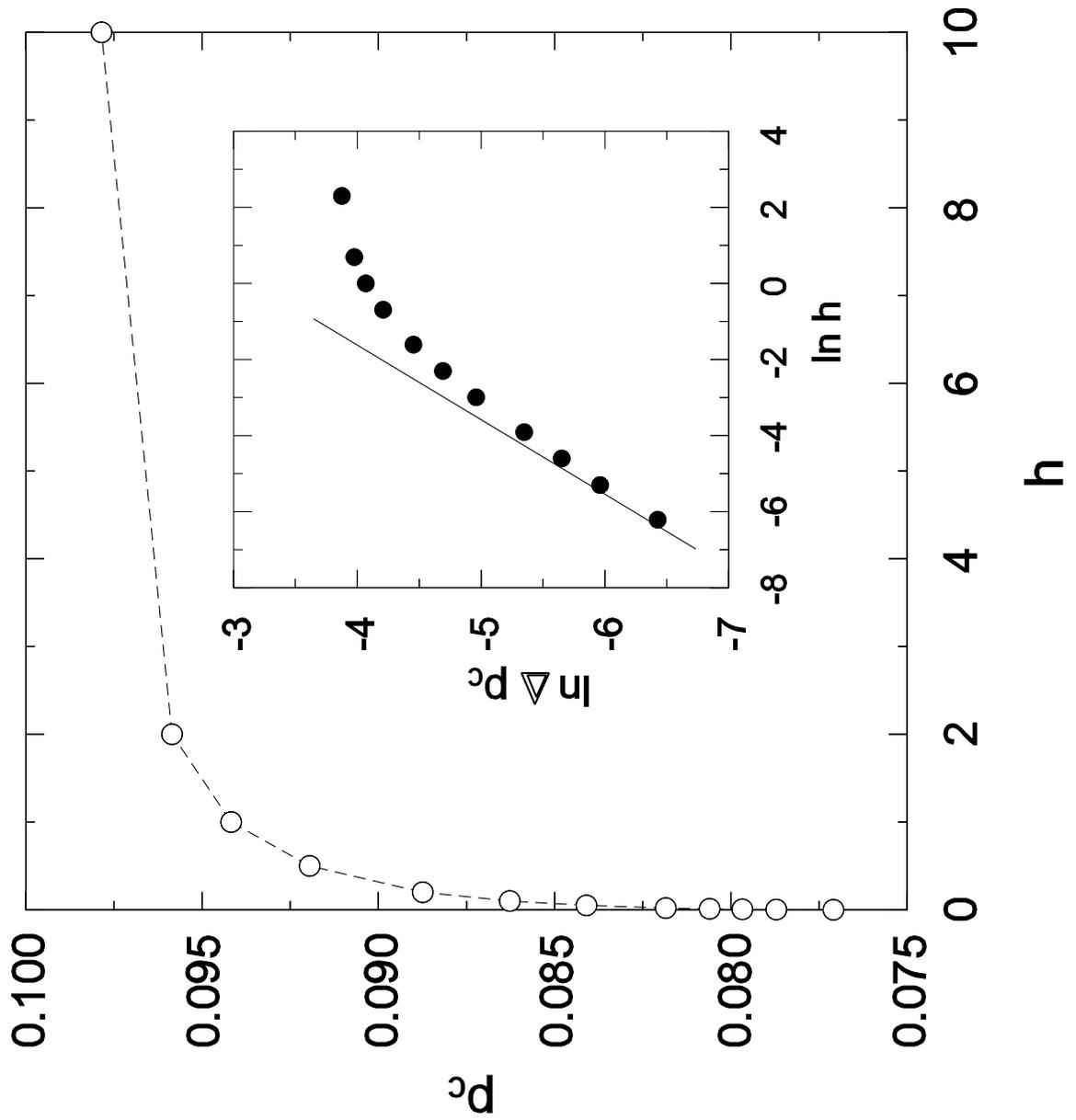
\end{document}